\documentstyle[12pt]{article}
\newlength{\pubnumber} 

\catcode`\@=11
\@addtoreset{equation}{section}

\def\section{\@startsection{section}{1}{\z@}{3.5ex plus 1ex minus .2ex}
 {2.3ex plus .2ex}{\large\bf}}
\def\subsection{\@startsection{subsection}{2}{\z@}{2.3ex plus .2ex}
 {2.3ex plus .2ex}{\bf}}

\begin{document}

\begin{titlepage}
\samepage{
\setcounter{page}{1}
\rightline{IASSNS-HEP-97/92}
\rightline{\tt hep-th/9708016}
\rightline{August 1997}
\vfill
\begin{center}
   {\Large \bf From the Type~I String to M-Theory:\\
       A Continuous Connection\\}
\vfill
   {\large
    Julie D. Blum\footnote{
     E-mail address: julie@sns.ias.edu}
     $\,$and$\,$
      Keith R. Dienes\footnote{
     E-mail address: dienes@sns.ias.edu}
    \\}
\vspace{.12in}
 {\it  School of Natural Sciences, Institute for Advanced Study\\
  Olden Lane, Princeton, N.J.~~08540~ USA\\}
\end{center}
\vfill
\begin{abstract}
  {\rm
    It is well-known that the $SO(32)$ and $E_8\times E_8$ heterotic
    strings can be continuously connected to each other in nine dimensions.
    Since the strong-coupling duals of these theories are respectively
    the $SO(32)$ Type~I theory and M-theory compactified on a line segment,
    there should be a corresponding continuous connection between
    the Type~I string and M-theory.  In this paper, we give an explicit
    construction of this dual connecting theory.  Our construction
    also enables us to realize the $E_8\times E_8$ heterotic string
    as a D-string soliton of the Type~I theory.  This provides a useful
    alternative description of the D-brane bound states previously
    discussed from a Type~I$'$ point of view.
   }
\end{abstract}
\vfill
\smallskip}
\end{titlepage}

\setcounter{footnote}{0}

\def\beq{\begin{equation}}
\def\eeq{\end{equation}}
\def\beqn{\begin{eqnarray}}
\def\eeqn{\end{eqnarray}}
\def\sosixteen{{$SO(16)\times SO(16)$}}
\def\e8{{$E_8\times E_8$}}
\def\V#1{{\bf V_{#1}}}
\def\half{{\textstyle{1\over 2}}}
\def\ttwo{{\vartheta_2}}
\def\tthree{{\vartheta_3}}
\def\tfour{{\vartheta_4}}
\def\ttwob{{\overline{\vartheta}_2}}
\def\tthreeb{{\overline{\vartheta}_3}}
\def\tfourb{{\overline{\vartheta}_4}}
\def\etainv{{\overline{\eta}}}
\def\Str{{{\rm Str}\,}}
\def\bone{{\bf 1}}
\def\chibar{{\overline{\chi}}}
\def\Jbar{{\overline{J}}}
\def\qbar{{\overline{q}}}
\def\calO{{\cal O}}
\def\calE{{\cal E}}
\def\calT{{\cal T}}
\def\calM{{\cal M}}
\def\calF{{\cal F}}
\def\calY{{\cal Y}}
\def\rep#1{{\bf {#1}}}
\def\ie{{\it i.e.}\/}
\def\eg{{\it e.g.}\/}
\def\eleven{{(11)}}
\def\ten{{(10)}}
\def\nine{{(9)}}
\def\Ip{{\rm I'}}
\def\oneprime{{I$'$}}
\hyphenation{su-per-sym-met-ric non-su-per-sym-met-ric}
\hyphenation{space-time-super-sym-met-ric}
\hyphenation{mod-u-lar mod-u-lar--in-var-i-ant}


\def\inbar{\,\vrule height1.5ex width.4pt depth0pt}

\def\IC{\relax\hbox{$\inbar\kern-.3em{\rm C}$}}
\def\IQ{\relax\hbox{$\inbar\kern-.3em{\rm Q}$}}
\def\IR{\relax{\rm I\kern-.18em R}}
 \font\cmss=cmss10 \font\cmsss=cmss10 at 7pt
\def\IZ{\relax\ifmmode\mathchoice
 {\hbox{\cmss Z\kern-.4em Z}}{\hbox{\cmss Z\kern-.4em Z}}
 {\lower.9pt\hbox{\cmsss Z\kern-.4em Z}}
 {\lower1.2pt\hbox{\cmsss Z\kern-.4em Z}}\else{\cmss Z\kern-.4em Z}\fi}

\def\NPB#1#2#3{{\it Nucl.\ Phys.}\/ {\bf B#1} (19#2) #3}
\def\PLB#1#2#3{{\it Phys.\ Lett.}\/ {\bf B#1} (19#2) #3}
\def\PRD#1#2#3{{\it Phys.\ Rev.}\/ {\bf D#1} (19#2) #3}
\def\PRL#1#2#3{{\it Phys.\ Rev.\ Lett.}\/ {\bf #1} (19#2) #3}
\def\PRT#1#2#3{{\it Phys.\ Rep.}\/ {\bf#1} (19#2) #3}
\def\CMP#1#2#3{{\it Commun.\ Math.\ Phys.}\/ {\bf#1} (19#2) #3}
\def\MODA#1#2#3{{\it Mod.\ Phys.\ Lett.}\/ {\bf A#1} (19#2) #3}
\def\IJMP#1#2#3{{\it Int.\ J.\ Mod.\ Phys.}\/ {\bf A#1} (19#2) #3}
\def\NUVC#1#2#3{{\it Nuovo Cimento}\/ {\bf #1A} (#2) #3}
\def\etal{{\it et al.\/}}

\long\def\@caption#1[#2]#3{\par\addcontentsline{\csname
  ext@#1\endcsname}{#1}{\protect\numberline{\csname
  the#1\endcsname}{\ignorespaces #2}}\begingroup
    \small
    \@parboxrestore
    \@makecaption{\csname fnum@#1\endcsname}{\ignorespaces #3}\par
  \endgroup}
\catcode`@=12

\input epsf


\section{Introduction}
\setcounter{footnote}{0}

In ten dimensions, there are only two supersymmetric heterotic string
theories:  the $SO(32)$ theory, and the $E_8\times E_8$ theory.
These theories have, however, very different strong-coupling duals:
the $SO(32)$ theory is believed to be dual to the $SO(32)$ Type~I
theory \cite{W,PW}, while the \e8\ theory is believed to behave at strong
coupling as a certain eleven-dimensional theory (M-theory)
compactified on a line segment \cite{HW}.  This difference in
the dual theories is particularly striking when one considers
the fact that the $SO(32)$ and \e8\ heterotic strings are closely related
in nine dimensions.  In fact, there exist two
types of relations between these theories in nine dimensions.
The first is a {\it discrete}\/ $T$-duality relation:  when each
theory is compactified on a circle and is subjected to a Wilson line
that breaks its gauge symmetry
to \sosixteen, the resulting theories are $T$-dual to each other.
Second, there is also a {\it continuous}\/ relation between these
two theories in nine dimensions:
as first discussed in Refs.~\cite{NSW,Ginsparg},
they can be continuously connected to each other through
deformations of background fields.
Both of these connections imply that
nine-dimensional compactifications of M-theory
should be closely related to those of the Type~I string.

In the case of the $T$-duality relation,
this expectation is indeed borne out by the following well-known observation.
The \e8\ string, when compactified on a circle,
has a strong-coupling dual which can be identified as
M-theory compactified first on a circle
and then on a line segment.  However,
M-theory compactified on a circle yields the Type~IIA string,
and the Type~IIA theory compactified on a line segment is equivalent to
the $T$-dual of the Type~I theory, also known as the Type~\oneprime\
theory.
Thus, since the Type~I theory is the strong-coupling dual of the $SO(32)$
heterotic string, we see that the same $T$-duality relation that exists
between the $SO(32)$ and \e8\ heterotic strings in nine dimensions
also exists for their strong-coupling duals.

Until now, however, the second relationship has not been demonstrated.
Specifically, no {\it continuous}\/ nine-dimensional connection has been
given that relates the Type~I string to an M-theory compactification.
It is the purpose of this note to explicitly construct this continuous
connection.

Our approach will be as follows.
First, we shall explicitly construct a nine-dimensional heterotic
string model which continuously interpolates between the ten-dimensional
$SO(32)$ and \e8\ heterotic strings.
Our model will be parametrized by the radius $R$
of the compactified dimension:
as $R\to \infty$, our interpolating model will reproduce the ten-dimensional
supersymmetric $SO(32)$ heterotic string, while as $R\to 0$, our
interpolating model
will reproduce the \e8\ string.
Note that this supersymmetric interpolating model
is similar to the {\it non}\/-supersymmetric interpolating
models that have recently been used \cite{BDone,BDtwo} for deriving
the strong-coupling duals of {\it non}\/-supersymmetric heterotic strings.

\begin{figure}[htb]
\centerline{\epsfxsize 4.0 truein \epsfbox {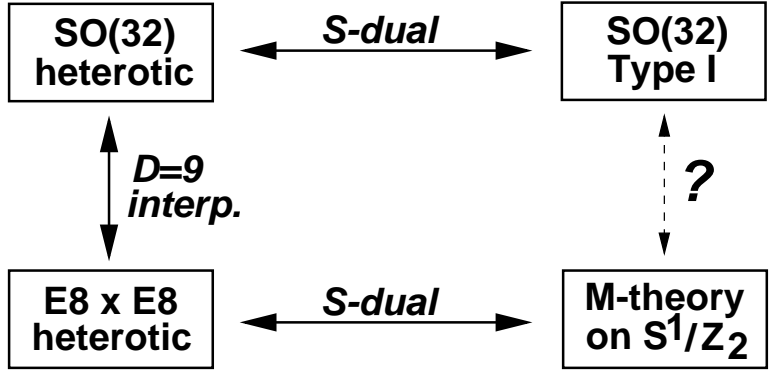}}
\caption{We construct a strong-coupling dual for the
    nine-dimensional heterotic string model that continuously interpolates
    between the ten-dimensional heterotic $SO(32)$ and $E_8\times E_8$
    string models.  This dual theory thereby provides a continuous
interpolation
    between the Type~I string and M-theory compactified on a line segment.
    Note that the $E_8\times E_8$ string is {\it equivalent}\/ to M-theory
    compactified on a line segment.
    In this figure,
    by contrast, the lower right box refers to the
    theory produced by compactifying M-theory
    first on a circle, then on a line segment, and then taking the
    radius of the circle to infinity.  This is equivalent to the
    Type~I$'$ theory at infinite coupling.}
\label{figone}
\end{figure}

Given this heterotic interpolating model,
we shall then proceed to construct its strong-coupling dual.
Our methods will be similar to those employed in Refs.~\cite{BDone,BDtwo}.
Thus, our dual theory will
in some sense interpolate between the Type~I theory and M-theory,
or more specifically between the Type~I theory and
the Type~I$'$ theory at infinite coupling.
This situation is illustrated in Fig.~\ref{figone}.

Once we have constructed this dual theory, we will examine
its D1-brane soliton.
We will find that this soliton reproduces the worldsheet
dynamics of the \e8\ heterotic string theory.
Note that because the \e8\ theory compactified
on a circle is related by strong/weak coupling duality to the
Type~I$'$ string,
it is expected that there exist D-particle bound
states in the Type~I$^\prime$ theory that give rise to the \e8\
gauge symmetry.
These states have been discussed in the context of the Type~I$^\prime$
theory in Ref.~\cite{DanFer}.
In this paper, by contrast, we will demonstrate that the D1-brane soliton
of our Type~I dual interpolating model gives rise to the expected
states.
Thus, our approach --- which is based entirely on the
Type~I point of view ---
provides the desired continuous connection
between the Type~I string and M-theory.

\section{Connecting $SO(32)$ to $E_8\times E_8$:
        The Heterotic Interpolating Model}
\setcounter{footnote}{0}

We begin by constructing our heterotic interpolating model
that continuously connects the ten-dimensional supersymmetric $SO(32)$
heterotic string and the ten-dimensional supersymmetric
\e8\ heterotic string.

To do this, we start with the ten-dimensional $SO(32)$ heterotic theory,
and we compactify this theory on a circle of radius $R_H$.
Next, we orbifold the resulting nine-dimensional theory by
the $\IZ_2$ element $Q'_H\equiv{\cal T} Q_H$ defined as follows.
The operator $\cal T$ is a translation of the coordinate
$x_1$ of the circle by half of the circumference of the circle:
\beq
       {{\cal T}:\hskip .2cm x_1\rightarrow x_1+\pi R_H ~.}
\label{Tdef}
\eeq
The operator $Q_H$ is the generator of the
orbifold
that produces the ten-dimensional supersymmetric \e8\
theory
from the ten-dimensional supersymmetric $SO(32)$ theory.
Note that if we decompose the representations of the original gauge group
$SO(32)$
into those of $SO(16)\times SO(16)$,
then $Q_H$ acts with a minus sign on the vector representation and one
of the spinor representations of the first $SO(16)$ factor.

This construction yields a supersymmetric
nine-dimensional heterotic string model
whose partition partition is given by:
\beqn
    Z(\tau) &=&  ({\rm Im}\,\tau)^{-7/2}\,{1\over (\overline{\eta}\eta)^7}
     \, (\chibar_V-\chibar_S)\, \times
     \,\biggl\lbrace \calE_0   (\chi_I^2 + \chi_S^2)
     +\calE_{1/2}   (\chi_I \chi_S+\chi_S\chi_I) \nonumber\\
   && ~~~~~
   +\calO_0  (\chi_V^2 + \chi_C^2)
   +\calO_{1/2}  (\chi_V \chi_C+\chi_C\chi_V)
             \biggr\rbrace~.
\label{ZE}
\eeqn
Here $\chi_{I,V,S,C}$ are the affine level-one characters of
the left-moving $SO(16)$ gauge factors;
$\chibar_{I,V,S,C}$ are the affine level-one characters
of the right-moving $SO(8)$ Lorentz group;
and the circle-compactification
functions $\calE,\calO$ are defined as
\beqn
   \calE_0(\tau,R_H) &\equiv&
    (\etainv\eta)^{-1}\,
    \sum_{ {m\in 2\IZ}\atop {n\in \IZ} }
     \, \exp\left\lbrack
        2\pi i mn\tau_1 ~-~
     \pi \tau_2 (m^2\alpha'/R_H^2 + n^2 R_H^2/\alpha') \right\rbrack\nonumber\\
   \calE_{1/2}(\tau,R_H) &\equiv&
    (\etainv\eta)^{-1}\,
    \sum_{ {m\in 2\IZ}\atop {n\in \IZ+1/2} }
     \, \exp\left\lbrack
        2\pi i mn\tau_1 ~-~
     \pi \tau_2 (m^2\alpha'/R_H^2 + n^2 R_H^2/\alpha') \right\rbrack\nonumber\\
   \calO_0(\tau,R_H) &\equiv&
    (\etainv\eta)^{-1}\,
    \sum_{ {m\in 2\IZ+1}\atop {n\in \IZ} }
     \, \exp\left\lbrack
        2\pi i mn\tau_1 ~-~
     \pi \tau_2 (m^2\alpha'/R_H^2 + n^2 R_H^2/\alpha') \right\rbrack\nonumber\\
   \calO_{1/2}(\tau,R_H) &\equiv&
    (\etainv\eta)^{-1}\,
    \sum_{ {m\in 2\IZ+1}\atop {n\in \IZ+1/2} }
     \, \exp\left\lbrack
        2\pi i mn\tau_1 ~-~
     \pi \tau_2 (m^2\alpha'/R_H^2 + n^2 R_H^2/\alpha')\right\rbrack
{}~.\nonumber\\
     && ~
\eeqn
Here $m$ and $n$ are respectively the momentum- and winding-mode numbers
on the circle,
and $\tau_1\equiv {\rm Re}\,\tau$, $\tau_2\equiv {\rm Im}\,\tau$.

These circle-compactification functions have the limits
\beqn
     R_H\to \infty:&~~~~~~~~& \calE_0, \calO_0\to
       \left( R_H\over 2\sqrt{\alpha'}\right)\,
       (\sqrt{\tau_2}\,\etainv\eta)^{-1}\,;
                  ~~ \calE_{1/2}, \calO_{1/2}\to 0 \nonumber\\
     R_H\to 0:&~~~~~~~~& \calE_0, \calE_{1/2}\to
       \left( \sqrt{\alpha'}\over R_H \right)\,
           (\sqrt{\tau_2}\,\etainv\eta)^{-1}\,;
                  ~~ \calO_0, \calO_{1/2}\to 0 ~.
\label{limits}
\eeqn
Since $(\sqrt{\tau_2}\etainv\eta)^{-1} $ is the
partition function of a single uncompactified boson, we
see that the relations (\ref{limits})
permit us to obtain the partition functions of ten-dimensional
string models as the limits of those in nine dimensions
(with the divergent radius factors providing the
proper overall volume factors).
Specifically, using these relations,
we see that the partition function $Z$ in Eq.~(\ref{ZE})
has the limits
\beqn
     \lim_{R_H\to\infty} Z &\sim& (\sqrt{\tau_2} \overline{\eta}\eta)^{-8}\,
         (\overline{\chi}_V-\overline{\chi}_S)
         (\chi_I^2 + \chi_V^2 + \chi_S^2 + \chi_C^2)~=~
             Z_{SO(32)}\nonumber\\
     \lim_{R_H\to 0} Z &\sim& (\sqrt{\tau_2} \overline{\eta}\eta)^{-8}\,
         (\chibar_V-\chibar_S)~(\chi_I + \chi_S)^2 ~=~
         Z_{E_8\times E_8} ~.
\eeqn
This demonstrates that our nine-dimensional model smoothly interpolates
between the ten-dimensional $SO(32)$ and \e8\ heterotic theories.
For generic radius $0< R_H <\infty$, the massless states of this model
consist of spacetime vectors and spinors transforming in the adjoint
representation
of \sosixteen.  As $R_H\to \infty$, the $SO(32)$ gauge symmetry is restored
through the appearance of extra massless vectors and spinors
in the $(\rep{16},\rep{16})$ representation of \sosixteen,
and as $R_H\to 0$ the $E_8\times E_8$ gauge symmetry is obtained
through the appearance of extra vectors and spinors
in the $(\rep{128},\rep{1}) \oplus (\rep{1},\rep{128})$ representation.
There are no extra massless particles at any finite non-zero radius.

\section{The Dual Interpolating Model}
\setcounter{footnote}{0}

Let us now construct the strong-coupling dual for our
heterotic interpolating model.

Since the $SO(32)$ heterotic string is conjectured to be dual to the
$SO(32)$ Type~I string, our dual interpolating model should
smoothly connect to the $SO(32)$ Type~I string.  This suggests
that our dual theory should be a smooth interpolation away from the
$SO(32)$ Type~I theory, and should include this theory at one endpoint.
In order to study interpolations of Type~I theories, we shall first
consider interpolations of Type~II theories, for
we know that Type~I theories can be obtained from Type~II theories
as orientifolds.
In particular, since the $SO(32)$ Type~I theory can be realized
as the orientifold of the Type~IIB theory, we shall begin by considering
the nine-dimensional Type~II interpolation that connects to the ten-dimensional
Type~IIB theory at one endpoint.

It turns out that there indeed exists such a Type~II model in nine dimensions
which connects to the Type~IIB
string at one endpoint and which is supersymmetric for all radii.
This model was explicitly described in Ref.~\cite{BDtwo},
and is similar to one of the models considered in Ref.~\cite{DHS}.
This model can be realized by compactifying the Type~IIB theory
on a circle of radius $R_I$, and then orbifolding by the action
$\calT$  where $\calT$ is the half-rotation operator
given in Eq.~(\ref{Tdef}).
The partition function of the resulting model is
\beq
    Z(\tau) ~=~  ({\rm Im}\,\tau)^{-7/2}\,{1\over (\overline{\eta}\eta)^7}
          \, (\calE_0 +\calE_{1/2})\,(\chibar_V-\chibar_S)\,(\chi_V-\chi_S)  ~
\label{ModelCprime}
\eeq
where we now use the level-one characters of $SO(8)$ for both the left-
and right-movers.

This model connects the supersymmetric Type~IIB theory at infinite radius
with the same model at zero radius.
However, as explained in Refs.~\cite{DHS,DLP}, the Type~IIB theory
at zero radius is equivalent to the Type~IIA
theory at infinite radius, and it is therefore the Type~IIA
theory that should be regarded as the effective ten-dimensional
theory that is produced in the $R_I\to 0$ limit.
Thus, our supersymmetric nine-dimensional
Type~II model interpolates between the ten-dimensional
Type~IIB and Type~IIA theories.

It is worth emphasizing that this is not the only nine-dimensional
Type~II model that interpolates between the Type~IIB and Type~IIA theories.
Another much more trivial example would be the straightforward
circle-compactification
of the Type~IIB theory, {\it without}\/ the orbifolding by $\calT$.  As the
radius of this circle goes to infinity, such a model would reproduce
the Type~IIB string;  it would also reproduce the Type~IIB string at zero
radius,
which is equivalent to the uncompactified Type~IIA string.  However, such
a trivial compactification would not be suitable for our purposes:
upon orientifolding, the resulting theory would not connect theories with
different gauge groups, and moreover the soliton of the resulting theory
would not have the desired properties.
Thus, as we shall see, the half-rotation orbifold element $\calT$ is crucial
not just on the heterotic side, but also on the dual side.

Given this Type~II interpolating model, we can now
construct its corresponding open-string orientifold.  We take
our orientifold generator to be $\Omega$, the worldsheet parity operator.
Thus, putting our entire construction together, we see that our open-string
interpolating model will be
a $\IZ_2\times \IZ_2$ orientifold of the supersymmetric Type~IIB
theory compactified on a circle of radius $R_I$.
The first $\IZ_2$ factor corresponds to the worldsheet parity operator
$\Omega$,
and the second to the orbifold operator $\calT$.
Because these two operators commute, we will be able to
consider the resulting open-string interpolating model either
as an orientifold of the Type~II interpolation,
or as an interpolating orbifold of the $SO(32)$ Type~I theory.
Of course, as $R_I\to\infty$, this open-string model will smoothly
reproduce the supersymmetric $SO(32)$ Type~I theory.

In order to construct this open-string theory,
we follow the standard orientifolding procedure \cite{orientifolds,DLP}.
Our conventions and notation will be the same as those of Ref.~\cite{BDtwo}.
We must evaluate eight traces, two each (corresponding to the NS-NS
and Ramond-Ramond sectors) for the torus, Klein bottle, cylinder,
and M\"obius strip.
Because this model is supersymmetric, the NS-NS and Ramond-Ramond
traces are equal to each other in all cases and provide cancelling
contributions.
Our traces are then as follows.
The total torus trace is simply half of the Type~II partition function given
in Eq.~(\ref{ModelCprime}).
Similarly, defining $q\equiv e^{-2\pi t}$,
$f_1(q)\equiv \eta(q^2)$, and $f_i(q)\equiv \sqrt{\vartheta_i(q^2)/\eta(q^2)}$
for $i=2,3,4$,
we find that
the NS-NS or Ramond-Ramond components
of the remaining traces are given as:
\beqn
       K'(t)  &=&
       {1\over 8}\, { f_4^8(q)\over f_1^8(q) } \,
         \left\lbrace \sum_{m=-\infty}^\infty\, [1+(-1)^m]\,q^{m^2 a^2/2}
       \right\rbrace~\nonumber\\
    C'(t) &=&
         {1\over 8}\, {f_4^8(q^{1/2}) \over f_1^8(q^{1/2})} \,
       \left\lbrace
        ({\rm Tr}\,\gamma_I )^2\,
              \sum_{m= -\infty}^\infty\, q^{m^2 a^2} ~+~
        ({\rm Tr}\,\gamma_\calT )^2\,
              \sum_{m= -\infty}^\infty\, (-1)^m \, q^{m^2 a^2}
        \right\rbrace\nonumber\\
    M'(t) &=&
         -\,{1\over 8}\,  \,{f_2^8(q) \, f_4^8(q) \over f_1^8(q) \, f_3^8(q)}
      \,\times\,\nonumber\\
        && ~ \left\lbrace
         ({\rm Tr}\,\gamma_\Omega^T \gamma_\Omega^{-1})\,
         \sum_{m= -\infty}^\infty \, q^{m^2 a^2} ~+~
         ({\rm Tr}\,\gamma_{\Omega \calT}^T \gamma_{\Omega \calT}^{-1})\,
         \sum_{m= -\infty}^\infty \, (-1)^m\, q^{m^2 a^2}  \right\rbrace ~.
\label{traces}
\eeqn
Here we have defined $a\equiv \sqrt{\alpha'}/R_I$;
the primes on $K'(t)$, $C'(t)$, and $M'(t)$ indicate that these traces do not
include
the integrations over non-compact momenta; and $\gamma_I$, $\gamma_\calT$,
and $\gamma_{\Omega \calT}$ respectively indicate the actions of the identity,
the orbifold element $\calT$, and the orientifold element $\Omega \calT$
on the Chan-Paton factors (or equivalently on the nine-branes in the
open-string theory).

In order to solve for these $\gamma$ matrices, we must impose the tadpole
anomaly cancellation constraints.
In general, tadpole anomalies appear as divergences in
the one-loop amplitudes from the $t\to 0$ region
of integration of these traces, and are interpreted as arising from the
exchange of massless (and possibly tachyonic) string states
in the tree channel.
In order to extract these divergences,
it is simplest to recast the above traces from the loop variable $t$
to the tree variable $\ell$, defined as
\beq
          \ell~\equiv~\cases{ 1/(4t) & Klein bottle\cr
                      1/(2t) & cylinder \cr
                      1/(8t) & M\"obius strip ~.\cr}
\label{tlconversion}
\eeq
The procedure for doing this is standard, and we find that
our above traces can be rewritten as
\beqn
       K'(\ell)  &=&
        {\ell^{-7/2}\over 64\,a}\,
       { f_2^8(\tilde q)\over f_1^8(\tilde q) } \,
         \sum_{m=-\infty}^\infty\, \left(
         \tilde q^{2 m^2/ a^2} ~+~ \tilde q^{ 2 (m+1/2)^2/ a^2} \right)
       \nonumber\\
    C'(\ell) &=&
        {\ell^{-7/2}\over 128 \, a}\,
       {f_2^8(\tilde q) \over f_1^8(\tilde q)} \,
       \left\lbrace
        ({\rm Tr}\,\gamma_I )^2\,
              \sum_{m= -\infty}^\infty\, \tilde q^{m^2/2 a^2} ~+~
        ({\rm Tr}\,\gamma_\calT )^2\,
              \sum_{m= -\infty}^\infty\, \tilde q^{(m+1/2)^2/2 a^2}
       \right\rbrace\nonumber\\
    M'(\ell) &=&
         -\,
        {\ell^{-7/2}\over 1024\,a}\,
       {f_2^8(\tilde q^{2}) \, f_4^8(\tilde q^{2}) \over f_1^8(\tilde
             q^{2}) \, f_3^8(\tilde q^{2})}
             \,\times\,\nonumber\\
        && ~ \left\lbrace
         ({\rm Tr}\,\gamma_\Omega^T \gamma_\Omega^{-1})\,
         \sum_{m= -\infty}^\infty \, \tilde q^{2 m^2/ a^2} ~+~
         ({\rm Tr}\,\gamma_{\Omega \calT}^T \gamma_{\Omega \calT}^{-1})\,
         \sum_{m= -\infty}^\infty \, \tilde q^{2 (m+1/2)^2/ a^2}
\right\rbrace~
\label{treetraces}
\eeqn
where $\tilde q\equiv e^{-2\pi \ell}$.
Taking the $\ell\to\infty$ limit,
and recalling that
the nine-dimensional Klein-bottle, cylinder, and M\"obius-strip
one-loop amplitudes $\int_0^\infty d\ell \,\ell^{7/2}\, \lbrace
K',C',T'\rbrace$
have relative prefactors of $512$,
$1$, and $512$ respectively,
we see that the NS-NS and Ramond-Ramond tadpoles will both be cancelled
for $R_I>0$ if
\beq
    64 ~+~ {1\over 16}\,({\rm Tr}\,\gamma_I)^2 ~-~ 4\,
     ({\rm Tr}\,\gamma_\Omega^T \gamma_\Omega^{-1})~=~ 0~.
\label{tadpole}
\eeq
These tadpoles cancel for any non-zero radius by choosing
$\gamma_\Omega$ symmetric and $\gamma_I$ to be 32-dimensional identity matrix.
Although we do not obtain a restriction on $\gamma_\calT$,
we can choose
\beq
     \gamma_\calT ~=~ \pmatrix{ \rep{1}_{16} & 0 \cr 0 & -\rep{1}_{16} \cr }~,
\label{Wilsonline}
\eeq
corresponding to the gauge group \sosixteen.
With this choice, we obtain open-string states consisting of a nine-dimensional
vector supermultiplet in the adjoint of \sosixteen.
When combined with the gravitational and Kaluza-Klein states from the
closed-string sector, this agrees exactly with the spectrum of massless states
of the heterotic interpolating model of Sect.~2.

Let us now discuss the $R_I\to 0$ limit.
In this case, we must be more careful when extracting the tadpole divergences.
In the $R_I\to 0$ limit,
the momentum sums in Eq.~(\ref{treetraces}) can be evaluated exactly:
\beq
     \lim_{a\to\infty} \,\sum_{m= -\infty}^\infty \,\tilde q^{A m^2/a^2}
     ~=~
     \lim_{a\to\infty} \,\sum_{m= -\infty}^\infty \,\tilde q^{A (m+1/2)^2/a^2}
     ~=~
      {a\over \sqrt{2\ell A}}
\eeq
for any normalization factor $A$.
Thus, as $\ell\to\infty$, the leading behavior of these traces is
\beqn
     K'(\ell) &\sim & (1/4) \,\ell^{-4}\, \lbrack 1 + ...\rbrack\nonumber\\
     C'(\ell) &\sim & (1/8) \,\ell^{-4}\,\lbrace
               ({\rm Tr}\, \gamma_I)^2 +
               ({\rm Tr}\, \gamma_\calT)^2
                    \rbrace \,\lbrack 1+...\rbrack \nonumber\\
     M'(\ell) &\sim & -(1/128)\,\ell^{-4}\,\lbrace
               ({\rm Tr}\, \gamma_\Omega^T \gamma_\Omega^{-1}) +
               ({\rm Tr}\, \gamma_{\Omega\calT}^T \gamma_{\Omega\calT}^{-1})
                    \rbrace \,\lbrack 1+...\rbrack ~.
\eeqn
The change in the power of $\ell$ reflects the change in the effective
dimensionality of the theory.
For such a ten-dimensional theory, the one-loop amplitudes now have the form
$\int_0^\infty d\ell\, \ell^4 \,\lbrace K',C',M'\rbrace$ with relative
prefactors $1024$, $1$, and $1024$ respectively.
We thus obtain the tadpole cancellation constraint
\beq
    256 ~+~
               {1\over 8}\,\lbrack ({\rm Tr}\, \gamma_I)^2 +
               ({\rm Tr}\, \gamma_\calT)^2 \rbrack
        ~-~
         8\,\lbrack
               ({\rm Tr}\, \gamma_\Omega^T \gamma_\Omega^{-1}) +
               ({\rm Tr}\, \gamma_{\Omega\calT}^T \gamma_{\Omega\calT}^{-1})
\rbrack ~=~ 0~.
\label{zeroradiustadpole}
\eeq
The only solution to this equation is one for which $\gamma_\calT$ is taken to
be the 32-dimensional identity matrix.  Unfortunately, the continuity of our
nine-dimensional Type~I interpolating model prevents us from making this
new choice for $\gamma_\calT$;  we must continue to have $\gamma_\calT$
as given in Eq.~(\ref{Wilsonline}).
However, given this choice, there is no solution to the
constraint equation (\ref{zeroradiustadpole}).
Specifically, this equation cannot be satisfied by adding or subtracting
nine-branes
(\ie, changing the rank of the gauge group), or even by adding anti-branes.
This is, of course, to be expected, for we know from ten-dimensional
anomaly cancellation arguments that there is no consistent
supersymmetric theory in ten dimensions with gauge group \sosixteen.

\section{The $E_8\times E_8$ Soliton}
\setcounter{footnote}{0}

The resolution to this ``puzzle'' is that there are extra non-perturbative
states
from the D1-brane soliton of this theory that become massless as $R_I\to 0$.
Let us therefore now discuss the soliton of this theory.
Our procedure for analyzing this soliton will be similar 
to the procedure followed in Ref.~\cite{BDtwo}, and consequently
we shall summarize here only the salient features.

Our open-string model contains a solitonic object or D1-brane.
Solitons of the $SO(32)$ Type~I theory corresponding to $SO(32)$ heterotic
strings
were found as classical solutions in Ref.~\cite{DabHull}, and were
constructed from a collective-coordinate expansion on
the D1-brane in Ref.~\cite{PW}.
In this section, we will construct
the soliton of our Type~I interpolating model by compactifying
the $SO(32)$ soliton on a circle and projecting
its worldsheet fields
by the orbifold element $\calY\equiv \calT\gamma_\calT$.

Note that the tension of the soliton is $T_S\sim T_F/\lambda_I^{(10)}$
where $T_F=1/2\pi\alpha'$
is the fundamental string tension and where $\lambda_I^{(10)}$ is the
ten-dimensional Type~I coupling.
If we were to choose the soliton to lie along any direction
orthogonal to the compact direction, the soliton would be very heavy for
all perturbative values of the coupling and would not behave as a fundamental
object.  Thus, we shall choose the soliton to lie along the compact $x_1$
direction because this choice allows states of the soliton to become nearly
massless
at perturbative values of the couplings for sufficiently small $R_I$.

In quantizing the fields on the soliton, we find that the quantized
zero-mode momentum of Type~I bosons in the $x_1$ direction becomes the
quantized oscillator moding of the worldsheet fields on the soliton.
Moreover, the Type~I theory
provides a restriction on the left- and right-moving
momenta of the form  $p_1^L=p_1^R$, and this restriction becomes the
worldsheet restriction $L_0=\overline{L_0}$
on the left- and right-moving Virasoro generators on the soliton.

Before the projection, the fields on the $SO(32)$ soliton are as follows.
The right-movers along $x_1$ consist of eight transverse
spatial bosonic fields $X^i$, $i=1,...,8$, and a Green-Schwarz fermion $S^a$,
$a=1,...,8$,
of definite chirality.
The left-movers consist of eight transverse spatial bosonic fields $X^i$
and 32 worldsheet fermions $\psi^A$, $A=1,...,32$.  There is also
a GSO projection acting on the left-moving fermions.

After we compactify this soliton on a circle,
we will project by the $\IZ_2$ element $\calY={\cal T}\gamma_\calT$.
Recall that $\gamma_\calT$ is the Type~I Wilson line that breaks
$SO(32)$ to $SO(16)\times SO(16)$.
It turns out that the action of $\gamma_\calT$ on the worldsheet
fields of the $SO(32)$ soliton exactly corresponds to the orbifold operator
$Q_H$ that we discussed in Sect.~2.  Since
$\gamma_\calT$ does not act on the Type~I spacetime, the spacetime action of
$\calY$
is solely due to $\cal T$ itself.  This
effectively halves the radius of the circle of compactification.
All worldsheet bosons of the soliton will thus be restricted to
integral moding such that these fields are periodic on the half-radius circle.

After the projection, there will be a variety of resulting sectors.
One group of sectors will
be those for which all fields have the usual modings on the half-radius circle
(such that the fields are periodic on the half-radius circle).
However, the reduction to the half-radius circle implies that there will
also be additional sectors for which some fields may be
anti-periodic on the half-radius circle (provided they are consistent with the
symmetries preserved by the orbifold).
Such unexpected modings on the half-radius circle arise naturally
when the original fields on the circle of radius $R_I$ are
decomposed into fields on the circle of half-radius $R_I/2$.
Taken together, then, all of these sectors will describe the Ramond sector of
the
resulting soliton, and just as for the supersymmetric $SO(32)$ soliton,
we will assume that a corresponding Neveu-Schwarz sector arises
non-perturbatively.

Because of the unbroken supersymmetry in our theory, we expect
that the massless states of our soliton will also be massless at
strong coupling.  Note that the soliton preserves half the supersymmetry
of the orientifold, resulting in a BPS condition for the soliton.
We will therefore analyze the soliton at weak coupling.  However, unlike the
case
of the non-supersymmetric solitons discussed in Ref.~\cite{BDtwo}, we will
project with $\calY$ separately on left- and right-movers, and in the end we
will allow combinations of left- and right-moving sectors with different types
of modings.

Performing the projection by $\calY$ is straightforward, and we can summarize
the results in the following table:
\beq
\begin{tabular}{||c|c||c|c||c|c||}
\hline
  \multicolumn{2}{||c||}{right-movers} & \multicolumn{4}{c||}{left-movers}
\\
\hline
\hline
  \multicolumn{2}{||c||}{~} & \multicolumn{2}{c||}{Ramond} &
\multicolumn{2}{c||}{NS} \\
\hline
   $\calY$ & sector & $\calY$ & sector &  $\calY$ & sector \\
\hline
 $+1$ & $\overline{V_8}$ & $+1$ & $S_{16}^{(1)}  S_{16}^{(2)}$
    &  $+1$ & $I_{16}^{(1)}  I_{16}^{(2)}$ \\
 $+1$ & $\overline{S_8}$ & $-1$ & $C_{16}^{(1)}  C_{16}^{(2)}$
    & $-1$ & $V_{16}^{(1)}  V_{16}^{(2)}$ \\
 ~ & ~
     & $+1$ & $I_{16}^{(1)}  V_{16}^{(2)}$
     & $+1$ & $S_{16}^{(1)}  C_{16}^{(2)}$ \\
 ~ & ~
     & $-1$ & $V_{16}^{(1)}  I_{16}^{(2)}$
     & $-1$ & $C_{16}^{(1)}  S_{16}^{(2)}$ \\
 ~ & ~
     & $+1$ & $I_{16}^{(1)}  S_{16}^{(2)}$
     & $+1$ & $S_{16}^{(1)}  I_{16}^{(2)}$ \\
 ~ & ~
     & $+1$ & $V_{16}^{(1)}  C_{16}^{(2)}$
     & $+1$ & $C_{16}^{(1)}  V_{16}^{(2)}$ \\
 ~ & ~
     & $+1$ & $S_{16}^{(1)}  V_{16}^{(2)}$
     & $-1$ & $V_{16}^{(1)}  S_{16}^{(2)}$ \\
 ~ & ~
     & $+1$ & $C_{16}^{(1)}  I_{16}^{(2)}$
     & $-1$ & $I_{16}^{(1)}  C_{16}^{(2)}$ \\
\hline
\end{tabular}
\label{solitontable}
\eeq
In this table, we have indicated the resulting
right- and left-moving sectors by specifying their representations under the
right-moving transverse Lorentz group $SO(8)$
and the two left-moving $SO(16)$ gauge factors respectively.
We also indicated their corresponding eigenvalues under $\calY$.

As our last step, in order to construct our final soliton theory, we must
join these left- and right-moving sectors together.
Our procedure for doing this is as follows.
First, as discussed above, we impose the requirement that $L_0=\overline{L_0}$.
Second,  we require that the exchange
symmetry of the two $SO(16)$ gauge factors be preserved.
Third, we require that all left/right combinations
be invariant under $\calY$.

Remarkably,
imposing these three restrictions simultaneously,
we find that we obtain precisely the eight sectors that comprise the
\e8\ heterotic string:
\beqn
     &&
     \overline{V_8}\, I_{16}^{(1)} I_{16}^{(2)} ~,~~
     \overline{V_8}\, S_{16}^{(1)} S_{16}^{(2)} ~,~~
     \overline{V_8}\, I_{16}^{(1)} S_{16}^{(2)} ~,~~
     \overline{V_8}\, S_{16}^{(1)} I_{16}^{(2)} ~\nonumber\\
     &&
     \overline{S_8}\, I_{16}^{(1)} I_{16}^{(2)} ~,~~
     \overline{S_8}\, S_{16}^{(1)} S_{16}^{(2)} ~,~~
     \overline{S_8}\, I_{16}^{(1)} S_{16}^{(2)} ~,~~
     \overline{S_8}\, S_{16}^{(1)} I_{16}^{(2)} ~.
\label{e8e8sectors}
\eeqn
Thus, we see that the \e8\ heterotic string can indeed be realized as
the D-string soliton of our Type~I interpolating model.
Moreover, we see from this analysis
that at $R_I=0$, the soliton provides
massless spacetime vectors and spinors in 
the $(\rep{128},\rep{1}) \oplus (\rep{1},\rep{128})$ 
representation of \sosixteen.
These additional states
thus enhance the gauge group
of our orientifold theory to \e8, and thereby cancel the anomaly
that we found earlier.

\section{Analysis and Discussion}
\setcounter{footnote}{0}

Let us now analyze our dual theory as a function of its radii and couplings.
This will be necessary in order for us to properly interpret our results.
Specifically, we seek to know the conditions under which
our \e8\ soliton indeed becomes massless at $R_I=0$.

We begin by recalling some basic relations.
If we compactify M-theory first on a line segment of
radius (length) $R_H^\eleven$ and then on a circle of
radius $R_H^\ten$, the resulting theory is equivalent
to the ten-dimensional \e8\ heterotic string with coupling
$\lambda_H$ compactified on a circle
of radius $R'_H$, where
\beq
   R_H^\eleven ~=~ \lambda_H^{2/3}~,~~~~~~
   R_H^\ten ~=~ { R'_H\over \lambda_H^{1/3} }~.
\eeq
Here $R_H^{(10,11)}$ are given in M-theory units, while $R'_H$ is given
in the usual string units.
Note that the prime on $R'_H$
is present in order to distinguish this radius variable
from $R_H$, the radius of our interpolating heterotic model in Sect.~2.
Specifically, our interpolating model at $R_H=0$ is $T$-dual to
the present theory at $R'_H=\infty$.

The strong-coupling dual of this compactification of M-theory can
be realized by exchanging the order of the compactifications:
first we compactify M-theory on a circle of
radius $R_\Ip^\eleven$ and then on a line segment of radius (length)
$R_\Ip^\ten$.
This produces the Type~I$'$ theory compactified on a circle of radius $R_\Ip$
with coupling $\lambda_\Ip$, where
\beq
   R_\Ip^\eleven ~=~ \lambda_\Ip^{2/3}~,~~~~~~
   R_\Ip^\ten ~=~ { R_\Ip\over \lambda_\Ip^{1/3} }~.
\eeq
Once again,
$R_I^{(10,11)}$ are given in M-theory units, while $R_I$ is given
in the usual string units.
Of course, these two theories are strong-coupling duals of each other only if
$R_H^\eleven= R_\Ip^\ten$ and $R_H^\ten=R_\Ip^\eleven$.
Finally, in order to realize the Type~I theory, we can $T$-dualize the
Type~I$'$ theory.  This produces the Type~I theory formulated at radius $R_I$
and coupling $\lambda_I$,
where
\beq
        R_I ~=~ {1\over R_\Ip}~,~~~~~~
        \lambda_I ~=~ {\lambda_\Ip \over R_\Ip}~.
\eeq

Let us now consider the mass of the soliton.
In general, the soliton has mass $M_{\rm sol}= T_F R_I/\lambda_I$ where
$T_F=(2\pi\alpha')^{-1}$ is the fundamental string tension.
Using the above relations, we can recast this expression in terms of
the fundamental heterotic variables $(R'_H,\lambda_H)$, obtaining
\beq
      M_{\rm sol} ~=~ {\lambda_H^{1/2}\over (R'_H)^{3/2}}~.
\eeq
Moreover, if our results are to be consistent,
we must have $M_{\rm sol}\to 0$ {\it at the same time}\/ that $R_I\to 0$.
Since
\beq
    R_I ~=~ {1\over (R'_H)^{1/2} \lambda_H^{1/2}}~,
\eeq
we see that these two equations do not have simultaneous solutions unless
$R'_H\to \infty$.
This is equivalent to $R_H\to 0$ for our interpolating model, which agrees
with our expectations on the heterotic side.
We are then permitted to have $\lambda_H\sim (R'_H)^\alpha$ where
$-1 < \alpha < 3$.
Thus, although $R'_H\to \infty$, we see that $\lambda_H$ is not constrained.
Note that these constraints imply that the Type~I$'$ coupling $\lambda_\Ip \to
\infty$.

Finally, let us now consider how our results compare with those of
Ref.~\cite{DanFer}.
In Ref.~\cite{DanFer}, the problem of realizing the \e8\ gauge
symmetry was considered from the Type~I$'$ point of view.
In such a picture, the problem reduces to showing the existence of
suitable bound states between mixtures of zero-branes and eight-branes.
In order to investigate the existence of such bound states, the authors of
Ref.~\cite{DanFer}
make a Born-Oppenheimer approximation which neglects
the non-abelian degrees of freedom involved in short-distance physics.
Note that the existence of certain
D0-bound states has been argued from a different perspective
in Ref.~\cite{sav}.
In this paper, by contrast, we have taken a Type~I approach
and have found a D1-brane soliton with exactly the worldsheet
fields of the \e8\ heterotic string.
This soliton satisfies a BPS condition,
so that its tension is not renormalized \cite{DHarv}.
Moreover, in the strong-coupling or small-radius limit,
our soliton can be expected to behave fundamentally as an \e8\ string,
and we know that
this theory is a stable solution to the string equations of motion.
Thus we have in some sense proven the existence of
all of the states necessary for correspondence with M-theory from the
Type~I point of view.

It is noteworthy that, just as for the heterotic model, there is exactly
one Type~I model that interpolates from $SO(32)$ gauge symmetry
at infinite radius to \e8\ gauge symmetry at zero radius.
Indeed, it is exactly this interpolation --- achieved through the use of
the half-rotation operator $\calT$ --- that enables us to realize
the \e8\ theory on the soliton.  We also see that at $R_I=0$,
the M-theory description is essential since
the Type~I$'$ theory is infinitely coupled
and ten-dimensional Lorentz
invariance is restored.
Thus,
our Type~I interpolating model succeeds in
continuously connecting the Type~I $SO(32)$ theory
to M-theory compactified on a line segment.

\bigskip
\medskip
\leftline{\large\bf Acknowledgments}
\medskip

We wish to thank K.~Intriligator and S.~Sethi for discussions.
This work was supported in part by
NSF Grant No.\  PHY-95-13835
and DOE Grant No.\ DE-FG-0290ER40542.

\vfill\eject
\bigskip
\medskip

\bibliographystyle{unsrt}

\end{document}